\documentclass[aps,pre,showpacs,floats,twocolumn,superscriptaddress,floatfix,footinbib]{revtex4-1}
\usepackage{graphicx}
\usepackage{amsmath,amsthm,amsfonts,amssymb,bm}
\usepackage{wasysym}
\usepackage{hyperref}
\usepackage[usenames,dvipsnames]{xcolor}
\usepackage{tikz}

\def\Bbb{\mathbb}

\newcommand{\beq}{\begin{equation}}
\newcommand{\eeq}{\end{equation}}
\newcommand{\beqa}{\begin{eqnarray}}
\newcommand{\eeqa}{\end{eqnarray}}
\newcommand{\beqan}{\begin{eqnarray*}}
\newcommand{\eenan}{\end{eqnarray*}}

\newcommand{\eqdef}{ {\kern 0.2em}={\kern -0.5em}:{\kern 0.2 em} }
\newcommand{\defeq}{ {\kern 0.2em}:{\kern -0.5em}={\kern 0.2 em} }

\newcommand{\Real}{{\mathbb R}}
\newcommand{\Nat}{{\mathbb N}}
\newcommand{\Int}{{\mathbb Z}}

\newcommand{\BZ}{\mathrm{BZ}}
\newcommand{\xfer}{{\Bbb T}}
\newcommand{\A}{{\Bbb A}}
\newcommand{\Id}{\mathrm{Id}}
\newcommand{\X}{{\cal X}}
\newcommand{\one}{\overline{1}}
\newcommand{\ip}[2]{\langle {#1} | {#2}\rangle}
\newcommand{\Num}{{\cal N}}

\newcommand{\finesse}{\delta}

\newcommand{\Vac}{{\varnothing}}

\newcommand{\setof}[2]{\left\{ #1 \middle| #2 \right\}}
\newcommand{\Lx}{C}
\newcommand{\Lt}{L}
\newcommand{\La}{{\cal B}_0}
\newcommand{\Lb}{{\cal B}_1}
\newcommand{\Ra}{{\cal T}_0}
\newcommand{\Rb}{{\cal T}_1}
\newcommand{\Latt}{\Int_C}

\begin{document}

\title{Triangular Ising antiferromagnet through a fermionic lens, part 2:
\\
information-theoretic aspects of zero-temperature states on cylinders}
 \author{Amir Nourhani}
 \email{nourhani@psu.edu}
 \affiliation{Department of Physics, Pennsylvania State University, University Park, PA 16802}
 \author{Vincent H. Crespi}
 \affiliation{Department of Physics, Pennsylvania State University, University Park, PA 16802}
 \affiliation{Department of Materials Science and Engineering, Pennsylvania State University, University Park, PA 16802}
 \affiliation{Department of Chemistry, Pennsylvania State University, University Park, PA 16802}
 \author{Paul E. Lammert}
 \email{lammert@psu.edu}
 \affiliation{Department of Physics, Pennsylvania State University, University Park, PA 16802}

\begin{abstract}

A classical lattice spin model wrapped on a cylinder is profitably viewed as a chain of rings of spins. From that perspective, mutual information between ring configurations plays much the same role as spin-spin correlation functions in simpler settings. We study zero-temperature states of triangular lattice Ising antiferromagnet (TIAFM) systems from this point of view using a fermionic representation presented in a companion paper (Part 1). On infinite cylinders, ring-to-ring mutual information falls off asymptotically at a rate which decreases smoothly with cylinder circumference, but the end-to-end mutual information for finite cylinders depends strongly on  the residue class modulo 3 of the circumference as well as on whether spin periodicity or antiperiodicity is imposed in the circumferential direction. In some cases, the falloff is only as the inverse square of the cylinder length. These features, puzzling within the original spin formulation, are easily understood and calculated within the fermionic formulation.
\end{abstract}

\maketitle 

\section{Introduction
\label{sec:intro}}

Traditionally, probabilistic dependence among the elementary degrees of 
freedom in statistical mechanical models is studied by means of 
correlation functions such as the two-spin correlator
for a model with Heisenberg spins.
This seems entirely natural on a regular lattice of any dimension. 
Consider, though, a spin model wrapped on a cylinder with a length 
much greater than its circumference. That is naturally viewed as
a one-dimensional system, the ``elementary'' constituents of which are rings of many
spins. Now, how do we measure the probabilistic dependence among these constituents?
Mutual information~\cite{Shannon+Weaver,Billingsley-ETI,Csiszar+Korner,Cover+Thomas},
which has been of increasing interest in classical statistical
mechanics~\cite{Lau+Grassberger-13,Wilms+-11,Melchert+Hartmann-13}, as well
as quantum information theory~\cite{Nielsen+Chuang-Book},
is one good answer.
Mutual information provides a precise quantification of the {\em amount\/} of 
dependence between two random variables of arbitrary complexity. 

In this paper, we apply this information-theoretic tool to the zero-temperature
triangular lattice Ising antiferromagnet (TIAFM) model on cylinders.
This model is an archetype of frustration ---
the presence of incompatible but equally strong elementary interactions ---
which occurs in an enormous range of systems, from 
water ice~\cite{Pauling-35,Giauque+Stout-36} to spin systems~\cite{Toulouse77,Moessner01,Normand09,Gingras14,Starykh15, Schmidt17}, artificial spin ice~\cite{nature1,Zhang+12,Perrin16}, colloidal assemblies~\cite{Tierno16, Han08}, Coulomb liquids~\cite{Mahmoudian15}, lattice gases~\cite{Weight03}, ferroelectrics~\cite{Choudhury11}, coupled lasers~\cite{Nixon13}, and self-assembled lattices of microscopic chemical reactors~\cite{Wang+Fraden-16}. 
Although the zero-temperature TIAFM is disordered, it has a short-distance rigidity in the form
of forbidden subconfigurations on individual triangles, as well as quasi-long-range
order in the infinite plane, as revealed by the power-law falloff of the spin-spin
correlator~\cite{Stephenson-64,Stephenson-70}.
This paper studies the effects of that rigidly and the vestiges of the
quasi-long-range order on cylinders by means of ring-to-ring and end-to-end
mutual information. 
Extensive investigations of the ground states of small TIAFM systems has been
performed by Millane and coworkers~\cite{Millane+Blakeley-04,Millane+Clare-06,Blakeley+Millane-06},
but information theoretic aspects have been so far neglected.
A central role in the current work is played by semi-conservation of particle number in 
the fermionic formulation of the TIAFM explained in the companion paper (Part 1) and a previous brief report~\cite{Nourhani+PRE-18a}.
This paper assumes Section 2 of Part 1 as background.
 
One way to read this paper is as a case study in the breakdown of the
``Perron-Frobenius scenario''. Transfer-matrix methods are a powerful, rigorous,
and simple way to understand the statistical mechanics of one-dimensional systems
with short-range interactions. At nonzero temperature, the transfer matrix for such
a system is aperiodic and irreducible. These are precisely the hypotheses of the
Perron-Frobenius theorem, which assures that the transfer matrix eigenvalue of 
largest modulus is nondegenerate and corresponds to an everywhere strictly positive
eigenvector. In turn, nondegeneracy of the largest eigenvalue implies the absence
of phase transitions and exponentially decaying influence of boundary conditions. 
The zero-temperature TIAFM on a cylinder violates the Perron-Frobenius scenario 
first because the transfer matrix is not irreducible. It connects a configuration
only to configurations of equal or smaller particle number, thus opening the door
to multiple phases. Further, for a circumference which is a multiples of three,
the eigenvalue of largest modulus can be nondegenerate, with a geometric multiplicity
of one, but an algebraic multiplicity of two. Among the consequences of that is
power law decay of end-to-end mutual information.

In the fermionic formulation of the TIAFM at
zero temperature, pairs of fermions can be annihilated as imaginary time 
passes, but not created. 
It will be necessary to fully incorporate that phenomenon into the formulation
in order to deal with effects of finite cylinder length quantitatively. 
Section \ref{sec:annihilation} handles that task, setting up crucial machinery 
which is applied in the balance of the paper.
To discuss the mutual information between the rings, we regard 
the cylindrical TIAFM as a one-dimensional lattice: 
\begin{center}
\tikz {\draw [-,dotted,thick] (-.13,0) -- (.13,0); \draw (0,0) ellipse (1.5pt and 6pt);}\!\!\!
\tikz {\draw [-,dotted,thick] (-.13,0) -- (.13,0); \draw (0,0) ellipse (1.5pt and 6pt);}\!\!\!
\tikz {\draw [-,dotted,thick] (-.13,0) -- (.13,0); \draw (0,0) ellipse (1.5pt and 6pt);}\!\!\!
\tikz {\draw [-,dotted,thick] (-.13,0) -- (.63,0); \draw (0,0) ellipse (1.5pt and 6pt);}\!\!\!
\tikz {\draw [-,dotted,thick] (-.13,0) -- (.13,0); \draw (0,0) ellipse (1.5pt and 6pt);}\!\!\!
\tikz {\draw [-,dotted,thick] (-.13,0) -- (.13,0); \draw (0,0) ellipse (1.5pt and 6pt);}\!\!\!
\tikz {\draw [-,dotted,thick] (-.13,0) -- (.13,0); \draw (0,0) ellipse (1.5pt and 6pt);}\!\!\!
\end{center}
in which each site of the chain is occupied by a ring of spins. 
The elementary degrees of freedom of the model are the spin configurations of 
entire rings, drawn from a configuration space $\X$ with $2^C$ elements. 
For brevity, we use the term {\it ring} also to refer to the spin configuration on a ring.
The cylindrical model is thus alternatively viewed as a chain of rings, each ring taking
a configuration from $\X$.
After a brief reminder of the idea of mutual information,
Section~\ref{sec:infinite-cylinders} calculates the 
ring-to-ring mutual information
on infinite cylinders, that is, the mutual information between the spin configurations
of distinct rings with particular attention to the asymptotic behavior as the
separation tends to infinity. The falloff rate of this quantity turns out
to be twice a spectral gap, that is, twice the rate of falloff of ordinary
correlation functions. A qualitative explanation of the difference is given.
Section~\ref{sec:finite-cylinders} turns to the end-to-end mutual information
on finite cylinders with free boundary conditions. The behavior in this case
turns out to be significantly more complicated, having non-monotonic dependence
on the circumference. Calculations paralleling those in 
Section \ref{sec:infinite-cylinders} are carried out and the origin of the
complicated behavior is exposed. The relevant energy gaps for finite cylinders
involve, in fermionic language, changing the particle number. In a more
statistical mechanical idiom, we may say that, asymptotically, the main 
channel of mutual information is global fluctuations between phases.

\section{particle nonconservation
\label{sec:annihilation}}
\begin{figure}
\includegraphics[width=90mm]{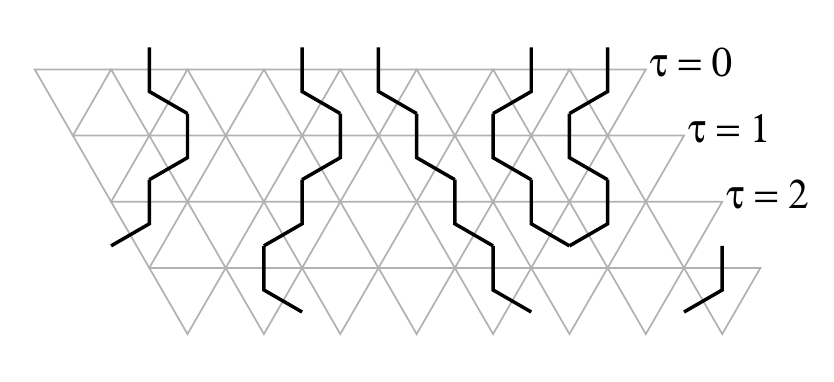}
\caption{
On this diagram of a cylindrical system, there are four strings;
one of them doubles back between times $2$ and $3$. 
In the fermionic picture, there are five fermions at time $0$. 
The doubling back is interpreted as annihilation of a pair of fermions.}
\label{fig:strings-recap}  
\end{figure}
This paper is concerned solely with zero-temperature TIAFM systems
on cylinders of finite circumference. Section 2 of Part 1 showed how
to represent bond configurations by string diagrams and, by interpreting
the strings as particle worldlines, reformulated the model as one of
fermions hopping on a ring and evolving in imaginary time. 
The number-conserving transfer matrix, obtained by 
forbidding the zero-temperature string diagram motif \includegraphics{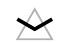},
was shown to be
\begin{equation}
\xfer_0 = e^{-{H}_0 + iP/2},  
\label{eq:T-exponential-form}
\end{equation}
where
\begin{equation}
{H}_0 = \sum_{q\in \BZ} {\varepsilon}(q) n(q),
\qquad
P = \sum_{q\in\BZ} q\, n(q),
\label{eq:H+P}
\end{equation}
are a Hamiltonian and total momentum operator, respectively.
$n(q) = c(q)^\dagger c(q)$ 
counts the number (0 or 1) of fermions in the mode of momentum $q$ and energy
\begin{equation}
{\varepsilon}(q) = -\ln\left(2\cos\frac{q}{2}\right).
\label{eq:e}
\end{equation}
The allowed fermion modes $q$ depend on the parity of the number of particles $\Num$,
and are given by
\begin{equation}
\label{eq:BZ}
\BZ = 
\begin{cases}
\frac{2\pi}{\Lx}\Int\cap (-\pi,\pi], & \Num \, \mathrm{odd}
\\
\frac{2\pi}{\Lx}\left(\Int+\frac{1}{2}\right)\cap (-\pi,\pi], & \Num\, \mathrm{even}.
\end{cases}
\end{equation}

The motif left out of the construction of $\xfer_0$, \includegraphics{annihilation_motif},
implements annihilation of neighboring pairs of fermions (see Fig.~\ref{fig:strings-recap}).
The consequences of that will now be considered in quantitative detail, via
two complementary approaches. Section \ref{sec:transfer-matrix-canonical}
incorporates pair-annihilation into the effective Hamiltonian, which thereby
becomes nonhermitian. The left- and right-eigenvectors of the transfer matrix
are found, making use of factorization over $(q,-q)$ subspaces.
Although this provides a complete analysis of $\xfer$, 
its interpretation in terms of the original statistical mechanical problem
is not so transparent.
Section \ref{sec:A-expansion} thus takes a different tack. Essentially, 
evaluation of the partition function of a cylindrical system is treated as a two-level
process. At the outer level, we sum over times of particle annihilation
events. Conditioned on those events, the partition sum is a product of shorter
cylinders with number-conserving boundary conditions which can be handled
as in Part 1.
Section \ref{sec:infinite-cylinder-states} applies the results to limiting 
states on infinite cylinders.

\subsection{full transfer matrix}
\label{sec:transfer-matrix-canonical}

A single pair annihilation event is depicted in Fig.~\ref{fig:strings-recap}.
The transfer matrix must ``decide'' how the configuration will evolve from
time $\tau$ to time $\tau+1$. This can be viewed as a two-stage process;
the first stage involves selecting pairs for annihilation, and the second,
moving what particles remain. 
This first stage is implemented by the operator
\begin{equation}
\label{eq:pair-destruction-0}
\xfer_{\mathrm{pr}} = \prod_{i\in\Latt} (1 + c_{i+1} c_i),
\end{equation}
which selects neighboring pairs in all possible ways from the
ring indexed by $\Latt$. Note that the operators in the product 
commute with each other, so it is unambiguous. 
Since the operators $\setof{c_{i+1}c_i}{i \in\Latt}$
square to zero, $1+c_{i+1} c_i = \exp(c_{i+1} c_i)$, and since they
commute with each other
$\xfer_{\mathrm{pr}} = \prod_{i\in\Latt} e^{c_{i+1} c_i}$. 
Applying commutativity again produces
$\xfer_{\mathrm{pr}} = e^{-H_{\mathrm{pr}}}$, where
\begin{equation}
\label{eq:T-pair-destruction}
H_{\mathrm{pr}} = 
-\sum_{i \in\Latt} 
 c_i c_{i+1} = \sum_{0 < q\in\BZ} 2i(\sin q)\,  c(-q) c(q).
\end{equation}
The exponentiated expression $e^{-H_{\mathrm{pr}}}$ can be re-expanded to
\begin{equation}
\xfer_{\mathrm{pr}}  = \prod_{q\in\BZ}[ 1 + 2i\sin q\,  c(q) c(-q)],
\end{equation}
a form we will use below.
The complete transfer matrix is then
\begin{equation}
\label{eq:T-two-exps}
\xfer = e^{-H_0} e^{-H_{\mathrm{pr}}}. 
\end{equation}

$\xfer_0$ conserves particle number $\Num$, so it can be written 
as $\xfer_0 =\bigoplus_{\Num} \xfer_{0,\Num}$, the direct sum of transfer
operators acting on $\Num$-eigenspaces.
 In fact, since $\xfer_0$ conserves
the number of particles (zero or one) in each $q$-state,
it can also be factorized over the individual $q$-spaces.
For ease of incorporating $\xfer_{\mathrm{pr}}$, however, we consider
a slightly coarser factorization of the fermion Fock space
${\cal F} = \bigotimes_{0\le q \in \BZ} {\cal H}(q)$,
where, labelling states by the occupied modes, 
${\cal H}(q)$ for $q > 0$ has basis
$|-q,q\rangle$, $|\Vac\rangle$, $|q\rangle$ and $|-q\rangle$, 
while ${\cal H}(0)$ has basis $|0\rangle$, $|\Vac\rangle$.
Correspondingly, $\xfer_0 = \bigotimes_{0\le q\in \BZ}\xfer_0(q)$,
with $\xfer_0(q)$ diagonal in the specified basis.
The full transfer matrix $\xfer = \xfer_0\xfer_{\mathrm{pr}}$ also factorizes
in the same way as
\begin{equation}
\xfer = \bigotimes_{0 \le q \in\BZ} {\xfer}(q).
\label{eq:T-factorized}
\end{equation}
Consider a generic nonzero $q$. 
${\xfer}(q) = \xfer_{\mathrm{e}}(q) \oplus \xfer_{\mathrm{o}}(q)$, where
$\xfer_{\mathrm{o}}(q)$ (subscript `o' for `odd') acts on the singly-occupied
subspace spanned by $|q\rangle$, $|-q\rangle$ and is simply
\begin{equation}
\xfer_{\mathrm{o}}(q) = e^{-\varepsilon(q)}
\left( 
\begin{array}{cc}
e^{iq/2} & 0 \\ 0 & e^{-iq/2}
\end{array}
\right).
\end{equation}
$\xfer_{\mathrm{e}}(q)$ acts on the even-occupancy subspace.
With the abbreviation $g(q) := 2\sin q$ and suppressing $q$ dependence
of both $g$ and $\varepsilon$, it has matrix
\begin{align}
\label{eq:Te(q)-diagonalized}
\xfer_{\mathrm{e}}(q) 
& =
\left( \begin{array}{cc}
e^{-2\varepsilon} & 0 \\
ig & 1
\end{array}  
\right)
\nonumber \\
& =
e^{-2\varepsilon}
\left( \begin{array}{c}
1 \\ \frac{i g}{e^{-2\varepsilon} -1}
  \end{array}\right)
  \left( \begin{array}{cc}
1 & 0    
  \end{array}\right)
\nonumber \\
& \quad +
\left( \begin{array}{cc}
0 \\ 1 
  \end{array}\right)
  \left( \begin{array}{cc}
\frac{-i g}{e^{-2\varepsilon} -1} & 1
  \end{array}\right),
\,\mathrm{if}\,\, q \neq \frac{2\pi}{3}.
\nonumber
\end{align}
The second, diagonalized, form holds for $\varepsilon(q) \neq 0$.
The exceptional zero-energy single particle modes occur only for 
$\Lx\in 3\Int$
in the odd-$\Num$ sector. In that case, we have
\begin{equation}
\label{eq:zero-energy-T}
\xfer_{\mathrm{e}}\left(\frac{2\pi}{3}\right) = 
\left(\begin{array}{cc} 1 & 0 \\ i\sqrt{3} & 1 \end{array}\right).
\end{equation}

Putting those exceptional cases aside, $\xfer$ has diagonalization
\begin{equation}
\label{eq:T-diagonalized}
\xfer = \sum \lambda_i |e_i \rangle\langle \theta^i|,
\quad \langle \theta^i|e_j\rangle = \delta_{i}^i.
\end{equation}
Compare this to the corresponding expansion
\begin{equation}
\xfer_0 = \sum \lambda_i |\varphi_i \rangle\langle \varphi_i|
\nonumber
\end{equation}
 for $\xfer_0$.
An eigenvector $\varphi_i$ essentially amounts to just a list of
occupied $k$-modes. 
$e_i$ is obtained from $\varphi_i$ by adding terms 
with one or more $(q,-q)$ pairs removed,
while $\theta_i$ is obtained by adding terms with one or more $(q,-q)$
pairs added. So,
$\langle \varphi_i | \theta^i\rangle = \langle \varphi_i | e_i\rangle = 1$.
We refer to $\varphi_i$ as the {\it parent state\/} of $e_i$ and $\theta^i$.
Note that $\xfer$ has the same eigenvalues as $\xfer_0$.
Also, $\ip{a}{e_i} = 0$ unless $\Num(a) \le \Num(\varphi_i)$.
Similarly, $\ip{\theta^i}{a} = 0$ unless $\Num(a) \ge \Num(\varphi_i)$.

\subsection{$\A$-expansion}
\label{sec:A-expansion}

The diagonalized form (\ref{eq:T-diagonalized}) of the transfer matrix is powerful. 
The partition function of a length-$\Lt$ cylinder with fixed-configuration boundary 
condition $a$ at the top end and $b$ at the bottom, drawn from the ring-configuration 
space $\X$, is 
\begin{equation}
\label{eq:Z-via-SVD}
Z_\Lt(b|a) = \ip{b}{ \xfer^\Lt a} 
= \sum \lambda_i^\Lt \ip{b}{e_i} \ip{\theta^i}{a}.
\end{equation}
Nevertheless, an alternative method of handling $\xfer_{\mathrm{pr}}$,
which is arguably more probabilistic in spirit, provides a useful
complement. 
To develop it, decompose $\xfer_{\mathrm{pr}}$ as
\begin{equation}
\label{eq:T-pr-decomp}
\xfer_{\mathrm{pr}} = \Id + \sum_{n=1}^{\lfloor{\Lx/2}\rfloor} \A_{n}, 
\end{equation}
where
\begin{equation}
  \label{eq:T-2n}
\A_{n} = \sum_{i_k +1 < i_{k+1} : 1 \le k \le n } 
(c_{i_1} c_{{i_1}+1}) \cdots (c_{i_n} c_{{i_n}+1}) 
\end{equation}
simply removes $n$ nearest-neighbor particle pairs in all possible ways.
Then,
\begin{equation}
\label{eq:T-in-A-expansion}
\xfer^\tau = 
\sideset{}{'}\sum_{\underline{\tau},\underline{n}}
\xfer_0^{\tau_{k}} \A_{n_k} \cdots \xfer_0^{\tau_1}\A_{n_1} \xfer_0^{\tau_0}.
\end{equation}
The sum here is over all values of $k \ge 0$, $\tau_0,\ldots,\tau_k$
and $n_1,\ldots,n_k$, and the prime on the sum indicates the constraint 
$\tau_0+\cdots + \tau_k = \tau$.
This expansion is eminently manageable because there can be at most
$\lfloor \Lx/2 \rfloor$
$\A$'s inserted. 
For the partition function (\ref{eq:Z-via-SVD}), we now obtain
\begin{align}
\label{eq:Z-in-A-expansion}
Z_\Lt(b|a) 
&=
\sideset{}{'}\sum_{\underline{x},\underline{y},\underline{n},\underline{\Lt}}
Z^0_{\Lt_{k}}(b|a_k)
\ip{ a_k}{\A_{n_k}b_k} \cdots
\nonumber \\
& \qquad
\ip{a_2}{\A_{n_2}b_2} 
Z^0_{\Lt_1}(b_2|a_1)
\langle a_1|\A_{n_1}b_1\rangle Z^0_{\Lt_0}(b_1|a),
\nonumber 
  \end{align}
with $Z_\Lt^0$
the number-conserving partition function constructed from $\xfer_0$.
Note that, with the intermediate states $a_i$, $b_i$ being delta functions
on single configurations in $\X$, the factors $\langle a_m|\A_{n_m} b_m\rangle$
are either zero or one. 
Effective use of the expansion (\ref{eq:T-in-A-expansion}) for large 
 $\Lt$
hinges
on the fact that $\xfer_0$ is transitive on each fixed-particle-number
configuration space $\X_N \subset \X$.
For then, according to the Perron-Frobenius theorem, the leading eigenvalue 
of the restriction $\xfer_0|_{\X_N}$ is nondegenerate and has strictly positive 
weight on each configuration in $\X_N$. Thus, 
if $N = \Num(a) = \Num(b)$, then for large 
$\Lt$,
\begin{equation}
Z_\Lt^0(b|a) \sim 
\langle b| \varphi_{N,0}\rangle \lambda_{N,0}^\Lt \langle \varphi_{N,0}| a\rangle,
\nonumber
\end{equation}
where $\lambda_{N,0}$ is the largest eigenvalue of $\xfer_0|_{\X_N}$
and $\varphi_{N,0}$ is the corresponding eigenvector.
More generally
\begin{equation}
\label{eq:Zba-asymptotic}
Z_\Lt(b|a)  \sim   
c \Big( \mathrm{max}\, \{\lambda_{N^*,0}\,;\, \Num(b) \le N^* \le \Num(a)\} \Big)^\Lt.
\end{equation}
Using the expansion (\ref{eq:T-in-A-expansion}), the argument for this
is that the sum is dominated by configurations where the $\A$'s are
clustered at either end so as to allow most of the length of the
cylinder to have $N^*$ particles, where $N^*$ is the maximizing value
in (\ref{eq:Zba-asymptotic}). The only exception to this is if
the maximizing $N^*$ is not unique. This
occurs only in the odd-$\Num$ sector with
 $\Lx\in 3\Nat$. 
 In that case, if 
$\Num(b) \le 2\Lx/3 -1$ and $\Num(a) \ge 2\Lx/3 + 1$, 
there will be a transition between the two particle numbers
effected by destroying two particles in the zero-energy mode,
and it will occurs indifferently anywhere along the length of 
the cylinder. The result in that case is 
$Z_\Lt(b|a) \sim c \Lt \lambda_{2\Lx/3-1,0}^\Lt$.

The $\A$-expansion can be leveraged to draw conclusions
about the eigenfunctions of the full transfer matrix $\xfer$. 
An important example is that,
\begin{align}
\langle b | e_0 \rangle &\neq 0 \;\,\mathrm{iff}\;\, \Num(b) \le \Num^p_c(\Lx),
\qquad
\nonumber \\
\langle \theta^0 | b \rangle & \neq 0 \,\;\mathrm{iff}\;\, \Num_c^p(\Lx)  \le \Num(b),
\label{eq:nonzero-overlaps}
\end{align}
where 
\begin{equation}
  \label{eq:N-critical}
\Num_c^{\mathrm{even}}(\Lx) = 2 \left\lfloor \frac{\Lx}{3} +\frac{1}{2} \right\rfloor,
\quad
\Num_c^{\mathrm{odd}}(\Lx) = 2 \left\lfloor \frac{\Lx}{3} \right\rfloor +{1}
\end{equation}
are the numbers of single-particle modes of {\em non-negative\/}
energy in the even-$\Num$ and odd-$\Num$ sectors, respectively.
We give the argument for $\langle b | e_0\rangle$. The implication from left
to right has been established above. So, take $a$ such that 
$\Num(a) = \Num_c^p(\Lx)$.
It is clear from the Perron-Frobenius theorem that both 
$\langle a | e_0 \rangle$ and $\langle \theta^0 | a \rangle$ are nonzero
since $\xfer_0$ is transitive on $\X_{\Num{a}}$.
From the $\A$-expansion, if 
$\Num(b)\le \Num^p_c(\Lx)$,
we know that $Z_\Lt(b|a) \sim c \lambda_0^\Lt$.
But, from (\ref{eq:Z-via-SVD}), this is impossible unless 
$\langle b | e_0\rangle \neq 0$.
The argument for the other equivalence is analogous.

In a similar vein, writing
\begin{equation}
\one := \sum_{a\in\X} a
\label{eq:one}  
\end{equation}
for the state uniformly distributed over all
configurations (this corresponds to open boundary conditions),
both $\ip{ \one }{ e_0}$ and $\ip{ \theta^0 }{ \one }$ are nonzero. 
To see the first of these, choose configurations $a,b$ such
that $\ip{ b}{e_0}\ip{ \theta^0 }{a } \neq 0$.
Then, by positivity of $\xfer_0$, 
 $Z_\Lt(\one | a) \ge Z_\Lt(b|a) \sim \lambda_0^\Lt$,
which implies that $\ip{ \one }{ e_0} \neq 0$.
An extension of the same reasoning leads to a conclusion which will be
useful in Section \ref{sec:end-to-end-MI}. Namely, suppose that the
first excited state, $\varphi_1$, has two more particles than does $\varphi_0$.
Then, for any configuration with $\Num(b) = \Num(\varphi_1)$, $\ip{ b }{ e_0} = 0$,
$\ip{ b }{ e_1} \neq 0$. So, there are configurations $a$ with
$Z_\Lt(b|a) \sim \lambda_1^\Lt$. 
As before, by positivity of $\xfer$, we conclude
that $\ip{ \theta^1 }{ \one } \neq 0$. Note that this does {\em not}
prove that $\ip{ \one }{ e_1 } \neq 0$.

\subsection{Ground macrostates on an infinite cylinder}
\label{sec:infinite-cylinder-states}

Now we examine the structure of ground macrostates on an infinite
cylinder. These are limits of sequences of states for a cylinder with fixed 
$\Lx$
and
$-\Lt_1 \le \tau \le \Lt_2$  as $\Lt_1, \Lt_2 \to \infty$, 
possibly at different rates.
From the earlier discussion of the $\A$-expansion around (\ref{eq:Zba-asymptotic}),
we know that with fixed configurations $b$ and $a$ as boundary conditions, the
bulk settles into the state $\varphi_{N^*,0}$, where 
$N^*$ with $\Num(b) \le N^* \le \Num(a)$ is the
unique maximizer of $\lambda_{N^*,0}$, at least if there are no zero-energy modes. 
Thus, the probability of the ring configurations
$X_{t}$ and $X_s$ at imaginary times $t$ and $s$ ($\ge t$) being $x$ and $y$,
respectively, is
\begin{equation}
\langle \delta_x(X_t) \delta_y(X_s) \rangle = 
\langle \varphi_{N^*,0}|y \rangle 
\frac{ Z_{s-t}^0(y|x)}{\lambda_{N,0}^{s-t}} \ip{ x }{ \varphi_{N^*,0}}.
\nonumber  
\end{equation}
$Z^0$ can be used here rather than $Z$ since there is no change of $\Num$.
The connected correlation function 
\begin{align}
\langle \delta_x(X_t); \delta_y(X_s) \rangle_\Num := &
\langle \delta_x(X_t) \delta_y(X_s) \rangle_\Num  
\nonumber \\
& - \langle \delta_x(X_t) \rangle_\Num \langle \delta_y(X_s) \rangle_\Num,
\end{align}
is thus
\begin{align}
\langle \delta_x(X_t) &; \delta_y(X_s) \rangle_\Num  
= 
\\
&\sum_{i>0}
\left(\frac{\lambda_{\Num,i}}{\lambda_{\Num,0}}\right)^{s-t}
\ip{ \varphi_{\Num,0}}{y}  \ip{y}{ \varphi_{\Num,i} }
\ip{ \varphi_{\Num,i} }{x} \ip{x}{ \varphi_{\Num,0} }.
\nonumber  
\end{align}
Generically, for functions $f,g:\X \rightarrow \Real$,
\begin{equation}
\label{eq:connected-correlation-general}
\langle f(X_t)\,;\, g(X_s) \rangle_\Num  
\sim c \left| \frac{ \lambda_{\Num,1} }{ \lambda_{\Num,0} }\right|^{s-t}.
\end{equation}

For odd $\Num$ and $\Lx\in 3\Int$
zero-energy modes complicate matters slightly.
The interesting boundary configurations have $\Num(a)$ big enough
to include the zero-energy modes and $\Num(b)$ small enough to exclude them.
Then, the total entropy of the entire system is insensitive to the location
of the transition (or ``domain wall''), and the distribution of its location is 
uniform over the entire length of the cylinder. With $s$ and $t$ as above fixed and finite,
the probability that domain wall occurs between them goes to zero. The state in the bulk
thus converges to a probabilistic mixture of $\varphi_{\Num,0}$ and $\varphi_{\Num+2,0}$,
where the former excludes the zero-energy modes and the latter includes them. 
The probability that $s$ and $t$ are on one side or the other of the domain wall 
is determined by the relative rates of divergence of 
 $\Lt_1$ and $\Lt_2$.
The limiting state is the same as that obtained by a mixture of number-conserving
boundary conditions.
Thus, in the infinite-length limit, the Perron-Frobenius scenario is effectively
restored, as far as the bulk is concerned. 
Every infinite-cylinder state is a probabilistic mixture of the extremal states
(pure phases) which have well-defined values of $\Num$.
Decay of connected correlation functions, as in Eq. (\ref{eq:connected-correlation-general}),
has the same relation to eigenvalues of the transfer matrix as for thermally disordered
systems.  

\section{ring-to-ring mutual information on infinite cylinders}
\label{sec:infinite-cylinders}

Traditional one- and two-spin
correlation functions, e.g., $\langle \sigma_i\rangle$ and $\langle \sigma_i\sigma_j\rangle$,
tell us everything there is to know about the distribution of one spin conditioned
on the value of another, not only how much the conditional
distribution differs from the marginal, but also in what
way (is the correlation ferromagnetic or antiferromagnetic?).
For more complex random variables, e.g. bond values on a ring
around the cylinder, it is not clear how to describe in what way they are
dependent, but a simple measure of the strength of dependence is available in
the form of {\it mutual information}. 
We recall~\cite{Shannon+Weaver,Billingsley-ETI,Csiszar+Korner,Cover+Thomas} that
the entropy of a discrete random variable $Y$ is given by
\begin{equation}
H(Y) = -\sum_y P_Y(y) \ln  P_Y(y).
\label{eq:entropy}
\end{equation}
The conditional entropy given a second random variable $X$ by
\begin{equation}
H(Y|X) = -\sum_x P_X(x) \sum_y P_Y(y|X=x) \ln  P_Y(y|X=x),
\label{eq:conditional-entropy}
\end{equation}
and the mutual information between the two by
\begin{align}
\label{eq:mutual-info}
I(X\! :\! Y) 
&= \sum_{x,y} P_{XY}(x,y) \ln \frac{P_{XY}(x,y)}{P_X(x)P_Y(y)}  
\nonumber \\
&= H(X) + H(Y) - H(X,Y)
\nonumber \\
&= H(Y) - H(Y|X).
\end{align}
The last form here obscures that $I(X\!:\! Y) = I(Y\!:\! X)$, but
expresses clearly that $I(X\!:\! Y)$ measures the average amount of
information $X$ carries about $Y$ insofar as it is the amount by which
the uncertainty about $Y$ is reduced by the knowledge of the value of $X$.
Calculations below will mostly work from the first expression in
(\ref{eq:mutual-info}).

\subsection{ring-to-ring mutual information on infinite cylinders}
\label{sec:MI-infinite-cylinder}

As shown in Section \ref{sec:infinite-cylinder-states},
a zero-temperature pure phase of an infinite cylinder (ring indices in $\Int$) 
is labelled by a definite particle number $\Num$, and in such a state the joint probability 
of configurations $X_0$ and $X_\tau$ on rings $0$ and $\tau$, respectively, is
\begin{equation}
{\kern -1.0em} P_\Num(X_0=a \,\&\, X_\tau=b) = 
          \langle \varphi_{\Num,0}|b \rangle 
          \frac{Z_\tau(b|a)}{\lambda_{\Num,0}^\tau}
          \langle a|\varphi_{\Num,0}\rangle.
\label{eq:P-infinite-cylinder}
\end{equation}
Since only the $\Num$-particle configuration subspace $\X_\Num$ will be relevant in 
this subsection, `$\Num$' subscripts will be dropped until further notice, 
in order to simplify expressions.
The probabilities in (\ref{eq:P-infinite-cylinder})
provide the building blocks for the mutual information
between $X_0$ and $X_\tau$, which thus has expression ($\Num$ now implicit)
\begin{equation}
\label{eq:mutual-info-infinite}
I_\infty(X_0\! :\! X_\tau) = \sum_{a,b\in \X_\Num} 
          \langle \varphi_{0}|b\rangle 
          \langle a|\varphi_{0}\rangle 
          \frac{Z_\tau(b|a)}{\lambda_{0}^\tau}
         \ln \frac{\lambda_{0}^{-\tau}Z_\tau(b|a)}
                   {\langle \varphi_{0}|a\rangle \langle b|\varphi_{0}\rangle}.
\end{equation}
Asymptotically in $\tau$, both $\lambda_{0}^{-\tau}{Z_\tau(b|a)}$ 
and the argument of the logarithm in (\ref{eq:mutual-info-infinite}) tend to one.
Thus, the leading contribution to the mutual information comes from the logarithm. 
Precisely, recalling that the restricted transfer matrix is diagonalized by
states labelled with $\Num$ $k$-vectors, (\ref{eq:mutual-info-infinite}) can be
rewritten as
\begin{align}
I_\infty(X_0\! :\! X_\tau) 
= 
\sum_{a,b,j} &
          \langle a|\varphi_{0}\rangle \langle \varphi_{0}|b\rangle 
          \ip{ b}{\varphi_{j}} \ip{ \varphi_{j}}{ a}\times
\nonumber \\
          &\quad \left(\frac{\lambda_{j}}{\lambda_{0}}\right)^\tau 
            \ln W_{\tau}(b|a),
\label{eq:mutual-info-infinite-2}
\end{align}
where
\begin{equation}
\label{eq:W-def-0}
W_{\tau}(b|a) \defeq 1 + \sum_{i> 0} 
               \left(\frac{\lambda_{i}}{\lambda_{0}}\right)^\tau
                      \frac{ \ip{ b }{\varphi_{i}} \ip{ \varphi_{i}}{ a} }
                           { \ip{ b }{\varphi_{0}}\ip{ \varphi_{0}}{a} }.
\end{equation}
At first glance, it would appear that the asymptotic behavior of 
\hbox{$I_\infty(X_0\! :\! X_\tau)$} is found by taking $j=0$ and replacing 
$\ln W_{\tau}(b|a)$ by the $i=1$ term in the sum. 
This is wrong for two reasons: some surprising cancellations take
place, and there are two distinct eigenvalues with second largest modulus.
The fermion ground state is a filled Fermi sea, and the lowest energy
excitations are particle-hole pairs. There are two such, and they are
converted one into the other by reversing all momenta. 
Correspondingly, $\overline{\lambda_{1}} = \lambda_{2}$. 
Expanding $\ln W_{\tau}(b|a)$ in powers of the sum, and using that
$\ip{b}{\varphi_{0}}$ is real, the first-order term (in the logarithmic expansion) is
\begin{equation}
\sum_{j} \sum_{i>0} 
       \left(\frac{\lambda_{j}\lambda_{i}}{\lambda_{0}^2}\right)^\tau
\sum_{a,b} 
  \ip{ b}{\varphi_{j}} \ip{ \varphi_{j}}{ a} 
{ \ip{ b }{\varphi_{i}} \ip{ \varphi_{i}}{ a} }.
\nonumber
\end{equation}
Since $\setof{\Psi(X)}{ X\in \X_\Num}$ 
comprises an orthonormal basis for the $\Num$-particle Hilbert space, 
when $j=0$, the sums over $a$ and $b$ produce factors $\ip{\varphi_0}{\varphi_i}$.
Checking the nominally next-largest terms, $i=j=1$ and $i=j=2$ are also seen to
give vanishing contributions since, for example,
$\ip{b}{\varphi_{1}}\ip{b}{\varphi_{1}} 
= \ip{\varphi_{2}}{b}\ip{b}{\varphi_{1}}$,
which, summed over $b$ yields zero.
Finally, though, $j=1$, $i=2$ and $j=2$, $i=1$ give nonvanishing contributions.
The result is
$2 |\lambda_{1}/\lambda_{0}|^{2\tau}$.
However, we are not done, since the second-order term in the expansion of
the logarithm competes with this, when $j=0$.
Evaluating this using the same observations as in the preceding calculations,
we find a contribution $-|\lambda_{1}/\lambda_{0}|^{2\tau}$.
All together, up to relatively exponentially small corrections,
and restoring the `$\Num$' indices, we find
\begin{equation}
\label{eq:infinite-cylinder-MI}
I_{\infty,\Num}(X_0\! :\! X_\tau) 
\sim \left| \frac{\lambda_{\Num,1}}{\lambda_{\Num,0}} \right|^{2\tau}.
\end{equation}
A remarkable feature of this result is that the coefficient is exactly one.
In fact, it depends only on some simple spectral properties of the transfer
matrix, as a review of our computations reveals. For a real transfer matrix
with a nondegenerate eigenvalue of largest modulus (e.g., Perron-Frobenius scenario)
and all generalized eigenspaces of subleading eigenvalue modulus having 
equal geometric and algebraic degeneracy, the mutual information is asymptotically
as in (\ref{eq:infinite-cylinder-MI}), with a prefactor equal to half the total
dimension of all eigenspaces of subleading modulus.
Comparing to (\ref{eq:connected-correlation-general}), we see
that \hbox{$I_{\infty,\Num}(X_0\! :\! X_\tau)$} decays at a rate twice that of
connected correlations 
of the form $\langle f(X_0)\,;\, g(X_\tau)\rangle_{\Num}$.

\subsection{explanation of the exponent}
\label{sec:MI-heuristics}

In the Perron-Frobenius scenario, generic truncated correlation functions
fall off as
\begin{equation}
\langle f(X_u) g(X_{u+s}) \rangle  - \langle f(X_u) \rangle \langle g(X_{u+s}) \rangle  
\sim c \left|\frac{\lambda_{\Num,1}}{\lambda_{\Num,0}}\right|^s,
\label{eq:truncated PF}
\end{equation}
asymptotically as the separation $s$ along the cylinder tends to infinity.
As just demonstrated, $I_{\infty,\Num}(X_0\! :\! X_\tau)$ falls off at twice 
the rate of an ordinary correlation function. 
This may seem inconsistent, but one must remember that the mutual information
is not an ordinary correlation function. 

An explanation will now be given of why $I_{\infty,\Num}(X_0\! :\! X_\tau)$ has the
decay rate it does. It is really just a matter of the character of mutual information.
Thus, we consider now an arbitrary pair of discrete random variables $X$ and $Y$,
not necessarily connected to the TIAFM at all, thereby also reducing notational clutter.
Denote possible values of $X$ and $Y$ by greek ($\alpha$,...) and
latin ($i$,...) letters, respectively, so that we can write simply
$P(i|\alpha)$ instead of $P(Y=i|X=\alpha)$, etc. Abbreviate
\begin{equation}
\delta_\alpha P(i) = P(i|\alpha) - P(i),  
\label{eq:delta}
\end{equation}
the idea being that $X$ and $Y$ are very nearly independent, so that 
$\delta_\alpha P(i)$ is small. We shall expand in its powers of these.
First, however, note that
\begin{equation}
P(\alpha) \delta_\alpha P(i) = P(i,\alpha) - P(i)P(\alpha)   
\label{eq:is truncated}
\end{equation}
is just an ordinary truncated correlation function.

Using the new notation, the mutual information between $X$ and $Y$ is
\begin{align}
I(X\! :\! Y) = 
& \sum_\alpha P(\alpha) \sum_i 
\Big\{ [P(i) + \delta_\alpha P(i)] \ln [P(i) + \delta_\alpha P(i)]
\nonumber \\
&       - P(i)\ln P(i) \Big\}
\nonumber
\end{align}
Expand in powers of the $\delta_\cdot P(\cdot)$'s to obtain
\begin{align}
I(X\! :\! Y) = & 
\sum_{\alpha,i}
[1 +  \ln P(i)] P(\alpha) \delta_\alpha P(i)
\nonumber \\
& + \frac{1}{2} \sum_{\alpha,i}
P(\alpha) \frac{ [\delta_\alpha P(i)]^2}{P(i)} + \cdots
\nonumber
\end{align}
Outside the square brackets in the first sum is the truncated correlation 
function (\ref{eq:is truncated}). Thus, the sum over $\alpha$ will vanish,
leaving just.
\begin{equation}
I(X\! :\! Y) \sim 
\sum_{\alpha,i}
\frac{ [P(\alpha) \delta_\alpha P(i)]^2}{P(\alpha)P(i)}.
\nonumber
\end{equation}
Again, the truncated correlation function appears, now squared.
Now interpret in the context where $X$ and $Y$ are
$X_u$ and $X_{u+s}$ to conclude that $I_{\infty,\Num}(X_0\! :\! X_\tau)$ 
falls off at twice the rate of an ordinary correlation function.

Note that expansion in powers of $\delta_\alpha P(i)$ will be thwarted
if the event $Y=i$ requires $X=\alpha$. This is the situation faced
for the end-to-end MI of the zero-temperature cylindrical TIAFM.


\section{end-to-end mutual information on finite-length cylinders}
\label{sec:finite-cylinders}

\subsection{energy gaps}
\label{sec:correlation-lengths}

As just noted, the correlation length in the $\Num$ phase is
twice the decay length of the ring-to-ring mutual information.
In terms of the fermion picture, the correlation length is the reciprocal 
of $\Delta\varepsilon = \ln|\lambda_{\Num,0}/\lambda_{\Num,1}|$,
the fundamental energy gap of the $\Num$-particle subspace. 
This section takes a detailed look at 
the {\it unrestricted\/} fundamental energy gaps, since those are the 
relevant ones for 
the decay of end-to-end mutual information under open boundary conditions.

Recall that according to (\ref{eq:BZ}), allowed single-particle momenta 
satisfy
\begin{equation}
\frac{q}{\finesse} \in 
\begin{cases}
\Int, & \Num \, \mathrm{odd},
\\
\Int+\frac{1}{2}, & \Num\, \mathrm{even},
\end{cases}
\label{eq:momenta}
\end{equation}
where 
 $\finesse = 2\pi/\Lx$
is the finesse of the momentum spectrum.
For an infinite cylinder, the even/odd distinction between these cases is not so important.
$\varphi_{\Num,0}$ corresponds to a filled Fermi sea with $n$ particles, and 
$\varphi_{\Num,1}$ is a particle-hole excitation with a particle in the highest
occupied mode of $\varphi_{\Num,0}$ promoted to the lowest unoccupied mode.
For large $\Lx$
this gap is $\Delta\varepsilon \simeq v(q_F)\finesse = 2\pi v(q_F)/\Lx$,
where the band velocity
\begin{equation}
v(q) = \frac{1}{2}\tan\frac{q}{2}
\nonumber
\end{equation}
was derived in Section 2 of Part 1.

When unrestricted gaps are considered,
the situation is more intricate and the distinction between even-$\Num$ 
and odd-$\Num$ sectors matters, since not only differences between 
energies of single-particle modes are relevant, but also their exact values.
Recall that, in the absence of zero-energy modes, 
eigenstates of $\xfer$ are descended from those of $\xfer_0$, with the same eigenvalues.
Thus, it suffices here to analyze the parent states.
The ground state has all modes with 
$|q| < q_0 = 2\pi/3 = (\Lx/3)\finesse$
occupied.
Writing  $\Lx = 3p+m$,
with $p,m\in \Nat$, we get 
$q_0/\finesse = p + m/3$.
We now use a linearization of the dispersion relation around $q_0$ to determine the
fundamental gaps. Only a couple of cases will require attention to the curvature.
Then, the energy cost of a particle-hole excitation is always 
$v(q_0) \finesse \simeq \sqrt{3}\pi/\Lx$, 
as noted in the previous paragraph.
But we will see that, over $\X$, these are never the lowest-energy excitations.

Now specialize to odd $\Num$. According to Eq. (\ref{eq:momenta}), allowed momenta
are integers. If  $\Lx\equiv 1$
 (mod 3), then removing the particles from 
modes $q = \pm (q_0 -\finesse/3)$ has an energy cost of only 
$(2/3) v(q_0) \finesse$. Clearly, this is the lowest-energy excitation. 
The conclusion is reported as an entry in Table 
\ref{tab:energy-gaps}.
In case $\Lx \equiv 2$
(mod 3), the lowest energy excitation is obtained by 
adding particles to $\pm (q_0 + \finesse/3)$, with the same cost.
The case $m=0$ is very different, and different from anything in the even $\Num$ sector,
for this is the case with zero-energy modes. The two degenerate states of $\xfer_0$,
one with all levels filled up to and including the zero-energy modes, and the other
with the zero-energy modes empty, descend to the two-dimensional generalized ground
eigenspace of $\xfer$. As far as $\xfer_0$ is concerned, the gap is zero. 

Turn now to the even $\Num$ sector. If $\Lx \equiv 1$
(mod 3), the
the lowest excitation adds particles to $\pm (q_0 + \finesse/6)$,
and if $L \equiv 2$
(mod 3), removes particles from $\pm (q_0 - \finesse/6)$.
In either case, the cost is $(1/3)v(q_0)\finesse$.
In case  $L\equiv 0$
(mod 3), the pp, hh, and ph excitations are all degenerate
in the linearized approximation, with energy $v(q_0)\finesse$. 
However, since $\varepsilon(q)$ is convex, the hh excitation will be slightly
lower in energy.

Apart from  $L\equiv 0$
(mod 3) with even $\Num$, there is
one other specific cylinder with  $\Lx\ge 3$
for which the linearized approximation
leads to an incorrect conclusion.
That is  $\Lx=5$
for odd $\Num$. In this case, the hh excitation is actually exactly
degenerate with the pp excitation, as a straightforward computation shows.

\begin{table}[h]
\setlength{\tabcolsep}{6pt}
  \centering
\begin{tabular}{c|ccc|ccc}
\multicolumn{1}{c|}{$\Num$ parity} & \multicolumn{3}{c|}{ even }  & \multicolumn{3}{c}{ odd } \\
\hline
 $\Lx$ mod 3        
 & 0      & 1    & 2    & 0   & 1    & 2 \\
excitation type      
& hh     & pp   &  hh  &  hh & hh   & pp  \\
 $\Delta\varepsilon\cdot \Lx/(\pi\sqrt{3})$ 
&  1     & 1/3  & 1/3  & 0   & 2/3  & 2/3 
\end{tabular}
\caption{
\label{tab:energy-gaps}
fundamental energy gaps $\Delta\varepsilon$ of $\xfer_0$ and corresponding excitation types. 
`hh' and `pp' indicate excitations involving removal (addition) of two particles.
Energies are reported in units of 
$v(q_0)\delta = (\sqrt{3}/2)(2\pi/\Lx) = \pi\sqrt{3}/\Lx$,
according to a linearized approximation of $\varepsilon(q)$. 
Due to the strict convexity of $\varepsilon(q)$, energies reported for
hh (pp) excitations are overestimates (underestimates), though
the relative error goes to zero as 
$\Lx\to\infty$. 
}
\end{table}

\begin{figure}
\setlength{\unitlength}{1cm}
\includegraphics[width=80mm]{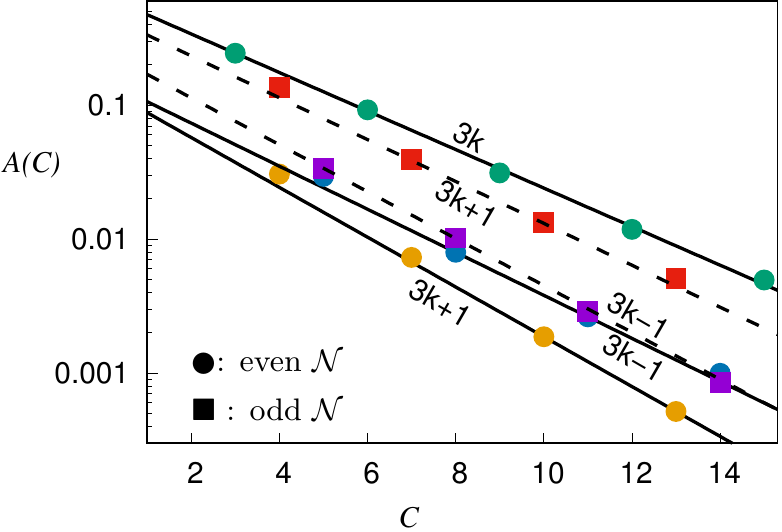}
\caption{Amplitudes $A(L)$ [see (\ref{eq:A-coeff-pp})] 
of the asymptotic decay of end-to-end mutual information~\cite{Nourhani+PRE-18a}.
}
\label{fig:end-to-end-mi-amp}  
\end{figure}
\begin{figure}
\setlength{\unitlength}{1cm}
\includegraphics[width=80mm]{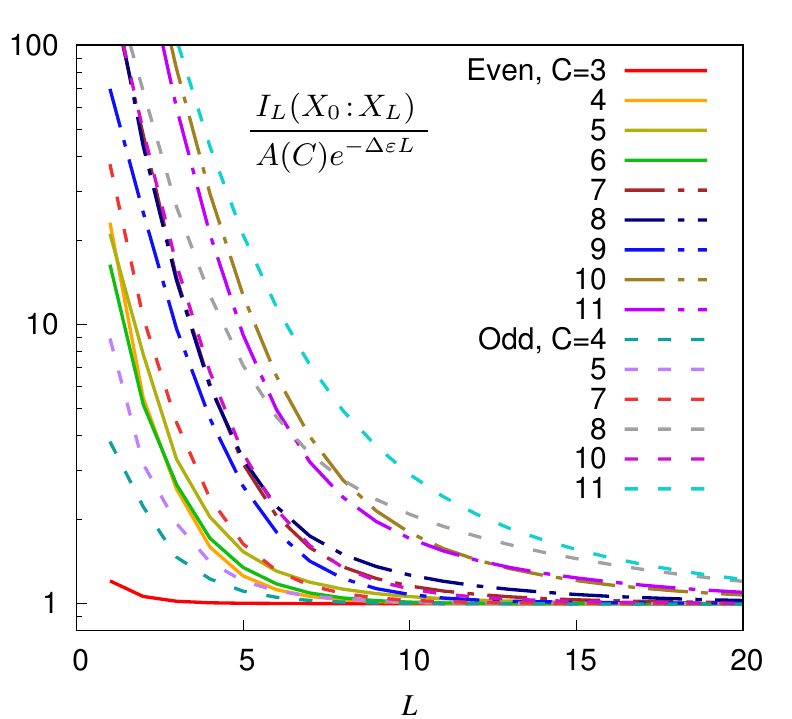}
\caption{Ratio of the end-to-end mutual information 
 \hbox{$I_\Lt(X_0\! :\! X_\Lt)$} 
to the leading behavior 
$A(\Lx) e^{-\Delta\varepsilon \Lt}$, 
for cases without zero-energy modes
and $3 \le \Lx \le 11$. The approach is always from above because
the leading behavior correctly accounts for all information transmission
via the first two eigenmodes, but not the others~\cite{Nourhani+PRE-18a}.
}
\label{fig:end-to-end-mi-scaled}  
\end{figure}

\subsection{open boundary end-to-end mutual information 
without zero-energy modes}
\label{sec:end-to-end-MI}

In this subsection, we compute the asymptotic behavior of the end-to-end
mutual information for a cylinder with open boundary conditions in cases
without zero-energy modes; the exceptional case 
$\Lx\equiv 0$ (mod 3),
$\Num$ odd,
will be treated in Section \ref{sec:MI-zero-energy-modes}.
The reader wishing to skip the intricacies of the calculations should go
to the results in 
(\ref{eq:end-to-end-mi-nonexceptional}),
(\ref{eq:A-coeff-pp}), and
(\ref{eq:A-coeff-hh}).

Mutual information is a property of a joint probability distribution.
In an equilibrium statistical mechanics context, probabilities come from
partition functions, which for the problem at hand just count ground 
microstates. 
As before, $Z_\Lt(b|a)$ here denotes the partition function 
with configurations $X_0$ and $X_\Lt$ on
the end rings of a length-$\Lt$ cylinder fixed to $a$ and $b$, respectively,
and $Z_\Lt(\one | \one)$ is the open boundary condition partition function.
The mutual information between the configurations $X_0$ and $X_\Lt$ at the
ends of a length-$\Lt$ cylinder is then
\begin{equation}
\label{eq:end-to-end-mi-Z}
I_\Lt(X_0\! :\! X_\Lt) = \sum_{(a,b)\in\X^2} 
                    \frac{Z_\Lt(b|a)}{Z_\Lt(\one | \one)}  
                    \ln\frac{Z_\Lt(b|a)Z_\Lt(\one | \one)}{Z_\Lt(b | \one)Z_\Lt(\one | a)}.
\end{equation}
$Z_\Lt(b|a)$ can be written in terms of the ring-to-ring transfer
matrix $\xfer$. 
In the absence of zero-energy modes,
\hbox{$\xfer = \sum_i \lambda_i |e_i\rangle\langle \theta^i|$} 
has eigenvalues $\lambda_0 > |\lambda_1| \ge |\lambda_2| \cdots$.
So, 
\begin{equation}
Z_\Lt(b|a) = \langle b|\xfer^\Lt a\rangle = 
\sum \lambda_i^\Lt \langle b | e_i\rangle \langle \theta^i | a\rangle.
\label{eq:Z_T(b|a)}
\end{equation}
To be able to extract the asymptotic behavior of 
\hbox{$I_\Lt(X_0\! :\! X_\Lt)$}, we
need to classify boundary configurations according to the largest-modulus eigenvalue
state with which they have nonzero overlap. 
Accordingly, we introduce the sets of configurations (${\cal T}$ for `top' and
${\cal B}$ for `bottom', i.e., of the cylinder)
\begin{align}
\La &\defeq \setof{b\in\X}{\ip{b}{\theta^0} \neq 0}, 
\nonumber \\
\Lb &\defeq
\setof{b\in\X}{\ip{b}{\theta^0} = 0, \ip{b}{\theta^1} \neq 0}, 
\nonumber \\
\Ra &\defeq \setof{a\in\X}{\ip{e_0}{a} \neq 0 }, 
\nonumber \\
\Rb &\defeq \setof{a\in\X}{ \ip{e_0}{a} = 0, \ip{e_1}{a} \neq 0}.
\end{align}
Various facts about the overlaps found near the end of Section \ref{sec:A-expansion}
will be called upon in the following. 

\subsubsection{$(b,a)\in \La\times \Ra$}

 If the contribution of these configurations does not vanish, then they dominate 
the asymptotic behavior of $I_\Lt(X_0\! :\! X_\Lt)$.
We shall see, however, that this contribution does vanish.
For these boundary configuration pairs, write
\begin{equation}
Z_\Lt(b|a) = \lambda_0^\Lt \langle b | e_0\rangle \langle \theta^0 | a\rangle  W_\Lt(b|a),
\end{equation}
with
\begin{equation}
  \label{eq:W-def}
W_\Lt(b|a) \defeq 1 + \sum_{i > 0} 
              \left(\frac{\lambda_i}{\lambda_0}\right)^\Lt
              \frac{\ip{b}{e_i} \ip{\theta^i}{a}}
                   {\ip{b}{e_0}\ip{\theta^0}{a}}.
\end{equation}
The contribution of $\La\times \Ra$ end configurations to the mutual information is
\begin{align}
\label{eq:end-to-end-MI-W}
I_\Lt^{\La\times \Ra}(X_0\! :\! X_\Lt) = \sum_{\La\times \Ra} \Big\{
      & \frac{\ip{b}{e_0}\ip{\theta^0}{a}}
           {\ip{\one}{e_0} \ip{\theta^0}{\one}}
                   \frac{W_\Lt(b|a)}{W_\Lt(\one\one)}  \times
\nonumber \\ &
\;                    \ln\frac{W_\Lt(b|a)W_\Lt(\one\one)}{W_\Lt(b\one)W_\Lt(\one a)}
\Big\}.
\end{align}
Asympototically as $\Lt\to \infty$,
\begin{equation}
\ln W_\Lt(b|a) \sim 
         \left(\frac{\lambda_1}{\lambda_0}\right)^\Lt
         \frac{\ip{ b }{ e_1 } \ip{ \theta^1 }{ a }}
              {\ip{ b }{ e_0 } \ip{ \theta^0 }{ a }}.
\label{eq:W-asymp}
\end{equation}
In writing this, we rely on the conclusion of the previous subsection
(Table \ref{tab:energy-gaps}), that for $\Lx \ge 3$ (with one exception: $\Lx=5$, odd $\Num$), 
there is a single nondegenerate eigenvalue of subleading modulus, $\lambda_1$, which is
real. Remarks on the exceptional case will be made momentarily.
Therefore,
\begin{widetext}
\begin{align}
\label{eq:end-to-end-mi-3}
I_\Lt^{\La\times \Ra}(X_0\! :\! X_\Lt) 
\sim 
\left(\frac{\lambda_1}{\lambda_0}\right)^\Lt
     \frac{1}{\ip{ \one }{ e_0 } \ip{ \theta^0 }{ \one }}
\sum_{b\in \La,a\in \Ra} 
\Bigg[ &
    \ip{ b }{ e_1 } \ip{ \one }{ a }
  +
    \ip{ b }{ e_0 } \ip{ \theta^0 }{ a }
      \frac{ \ip{ \one }{ e_1 } \ip{ \theta^1 }{ \one } }
           { \ip{ \one }{ e_0 } \ip{ \theta^0 }{ \one } }
\nonumber \\
&   -\ip{ b }{ e_1 } \ip{ \theta^0 }{ a }
      \frac{ \ip{ \theta^1 }{ \one } }
           { \ip{ \theta^0 }{ \one } }
  -
    \ip{ b }{ e_0 } \ip{ \theta^1 }{ a }
      \frac{ \ip{ \one }{ e_1 } }
           { \ip{ \one }{ e_0 } }
\Bigg].  
\end{align}
\end{widetext}
The sum here actually vanishes.
To see that, consider the two cases
$\Num(\varphi_1) = \Num(\varphi_0) - 2$, or
$\Num(\varphi_1) = \Num(\varphi_0) + 2$.
In the first case,
$\ip{b}{e_1} \neq 0$ implies that $\Num(b) \le \Num(\varphi_1) = \Num(\varphi) - 2$.
But, then, $\ip{b}{e_0} \neq 0$,
and the sum over $b$ can be expanded to all of $\X$ since the added terms
give zero anyway. The resulting sums over $b$ of the first term in square
brackets cancels that of the final term, and similarly the second and third
terms. 
In the other case, $\Num(\varphi_1) = \Num(\varphi_0) + 2$, proceed similarly,
but expand the range of $a$ instead. 
In the exceptional case ($\Lx=5$, odd $\Num$), expression (\ref{eq:W-asymp}) would
have a second term, but their separate contributions are zero, as just seen.

Thus, $I_\Lt^{\La\times \Ra}(X_0\! :\! X_\Lt)$ is exponentially small compared to 
$(\lambda_1/\lambda_0)^\Lt$; we must turn to the contributions of other boundary configurations:
$b$ in $\Lb$ or $a$ in $\Rb$.
The first case requires $\Num(b) \ge \Num(\varphi_1) = \Num(\varphi_0) +2$, 
hence essentially corresponds to a pp excitation,
whereas the second case requires
$\Num(a) \le \Num(\varphi_1) = \Num(\varphi_0) -2$, corresponding to a hh 
excitation. We compute their contributions separately and then add them.
Since the calculations are similar, only the $\Lb\times \Ra$ need be
discussed in any detail.
\subsubsection{$(b,a) \in \Lb\times \Ra$}
In this case,
\begin{equation}
\frac{Z_\Lt(b|a)}{Z_\Lt(\one | \one)} \sim 
         \frac{\ip{ b }{ e_1} \ip{ \theta^1 }{ a}}
              {\ip{ \one }{ e_0} \ip{ \theta^0 }{ \one }} 
\left(\frac{\lambda_1}{\lambda_0}\right)^\Lt, 
\nonumber
\end{equation}
and
\begin{equation}
\frac{Z_\Lt(b|a)Z_\Lt(\one | \one)}{Z_\Lt(b | \one)Z_\Lt(\one | a)} \sim 
         \frac{\ip{\theta^1}{a}\ip{\theta^0}{\one}}
              {\ip{\theta^1}{\one}\ip{\theta^0}{a}}.
\label{eq:end-to-end-10}
\end{equation}
Inserting these into (\ref{eq:end-to-end-mi-Z}), we find a contribution
to the end-to-end mutual information of
\begin{equation}
\label{eq:end-to-end-mi-nonexceptional}
I_\Lt^{\Lb\times\Ra}(X_0\! :\! X_\Lt) \sim 
A^{\Lb\times\Ra}(\Lx) \left(\frac{\lambda_1}{\lambda_0}\right)^\Lt,
\end{equation}
where 
\begin{align}
A^{\Lb\times\Ra}(\Lx) 
  = &\sum_{\Num(b) > \Num(\varphi_0) ,\, a\in \Ra}
         \frac{\ip{ b }{ e_1} \ip{ \theta^1 }{ a}}
              {\ip{ \one }{ e_0} \ip{ \theta^0 }{ \one }} 
        \ln \frac{\ip{ \theta^1 }{ a} \ip{ \theta^0 }{ \one }}
              {\ip{ \theta^1 }{ \one }\ip{ \theta^0 }{ a } }
\nonumber \\
  = &\sum_{\Num(a)\ge \Num(\varphi_0) +2}
         \frac{\ip{ \one }{ \varphi_1} \ip{ \theta^1 }{ a}}
              {\ip{ \one }{ e_0} \ip{ \theta^0 }{ \one }} 
        \ln \frac{\ip{ \theta^1 }{ a} \ip{ \theta^0 }{ \one }}
              {\ip{ \theta^1 }{ \one }\ip{ \theta^0 }{ a } }.
\label{eq:A-coeff-pp}
  \end{align}

\subsubsection{$(b,a) \in \La \times \Rb$}

This case is very similar to the previous one, so we omit the details.
The result is as in (\ref{eq:end-to-end-10}), with the analog of 
(\ref{eq:A-coeff-pp}) being
\begin{equation}
A^{\La\times\Rb}(\Lx) =  \sum_{\Num(b)\le \Num(\varphi_0)-2}
         \frac{\ip{ b }{ e_1} \ip{ \varphi_1 }{ \one}}
              {\ip{ \one }{ e_0} \ip{ \theta^0 }{ \one }} 
        \ln \frac{\ip{ b }{ e_1 } \ip{ \one  }{ e_0 }}
              {\ip{ \one }{ e_1 }\ip{ b }{ e_0 } }.
\label{eq:A-coeff-hh}
  \end{equation}

\subsubsection{final result}

Collecting the pp contribution (\ref{eq:A-coeff-pp}) from $\Lb \times \Ra$
boundary configurations and the hh contribution (\ref{eq:A-coeff-hh})
from $\La \times \Rb$ boundary configurations, the final result is
\begin{equation}
I_\Lt(X_0\! :\! X_\Lt) \sim 
A(\Lx) \left(\frac{\lambda_1}{\lambda_0}\right)^\Lt,
\end{equation}
with
\begin{align}
A(\Lx) 
&=  
 \sum_{\Num(a)\ge \Num(\varphi_0) +2}
         \frac{\ip{ \one }{ \varphi_1} \ip{ \theta^1 }{ a}}
              {\ip{ \one }{ e_0} \ip{ \theta^0 }{ \one }} 
        \ln \frac{\ip{ \theta^1 }{ a} \ip{ \theta^0 }{ \one }}
              {\ip{ \theta^1 }{ \one }\ip{ \theta^0 }{ a } }
\nonumber \\
&+ \sum_{\Num(b)\le \Num(\varphi_0)-2}
         \frac{\ip{ b }{ e_1} \ip{ \varphi_1 }{ \one}}
              {\ip{ \one }{ e_0} \ip{ \theta^0 }{ \one }} 
        \ln \frac{\ip{ b }{ e_1 } \ip{ \one  }{ e_0 }}
              {\ip{ \one }{ e_1 }\ip{ b }{ e_0 } }.
\end{align}
Using results of Section \ref{sec:transfer-matrix-canonical}, one can compute the required $e_0$, $\theta_0$, $e_1$ and $\theta_1$, and then the sums in (\ref{eq:A-coeff-pp},\ref{eq:A-coeff-hh}). Results for small values of $\Lx$ are plotted in Fig.~\ref{fig:end-to-end-mi-amp}. For each parity and residue class modulo 3, the amplitudes appear to  have very nearly an exponential dependence on $\Lx$. The reason for this is unclear.
Fig. \ref{fig:end-to-end-mi-scaled} shows the ratio of the asymptotic approximation $A(\Lx) e^{-\Delta\varepsilon \Lt}$ for the end-to-end mutual information $I_\Lt(X_0\! :\! X_\Lt)$ to the exact value, which is obtained from iterated powers of the transfer matrix working purely in position space. In the case of $\Lx=5$ with odd $\Num$, the sub-leading eigenvalue is degenerate and both (\ref{eq:A-coeff-pp}) 
and (\ref{eq:A-coeff-hh}) are counted, one for each eigenvector. 
\subsection{comparison to infinite cylinders}
\label{sec:MI-comparison}

Eq. (\ref{eq:end-to-end-mi-nonexceptional}) reveals a behavior of the end-to-end mutual information $({\lambda_1}/{\lambda_0})^\Lt$ very different from what was found for ring-to-ring mutual information in an infinite cylinder,  $|{\lambda_1}/{\lambda_0}|^{2\tau}$. They differ because a Taylor expansion in shifts 
(\ref{eq:delta}) is not valid for the important configurations, but more explanation of (\ref{eq:end-to-end-mi-nonexceptional}) can be given.

Defining $S_0$ to be $+$, $0$,or $-$ when $\Num(X_0)$ is greater than, equal to, or less than $\Num_c$, and similarly $S_\Lt$ from $X_\Lt$, two applications of the data-processing inequality~\cite{Cover+Thomas} yields
\begin{equation}
I(X_0:X_\Lt) \ge I(S_0:S_\Lt)  
\end{equation}
\begin{align}
I(S_0:S_\Lt)  &\ge P(S_0=-)[H(S_\Lt) - H(S_\Lt|S_0=-)]
\nonumber \\
& = P(S_0=-)H(S_\Lt).
\nonumber
\end{align}
Similarly,
\begin{equation}
I(S_0:S_\Lt)  \ge P(S_\Lt=+)H(S_0).
\nonumber
\end{equation}
Since one or the other of $P(S_\Lt=+)$ and $P(S_0=-)$ falls off as $(\lambda_{1}/\lambda_0)^\Lt$ (depending on whether the pp or hh excitation is lower energy), so will $I(S_0:S_\Lt)$.

\subsection{end-to-end mutual information with zero-energy modes}\label{sec:MI-zero-energy-modes}

\begin{figure}\centering
\setlength{\unitlength}{1cm}
\includegraphics[width=8.5cm]{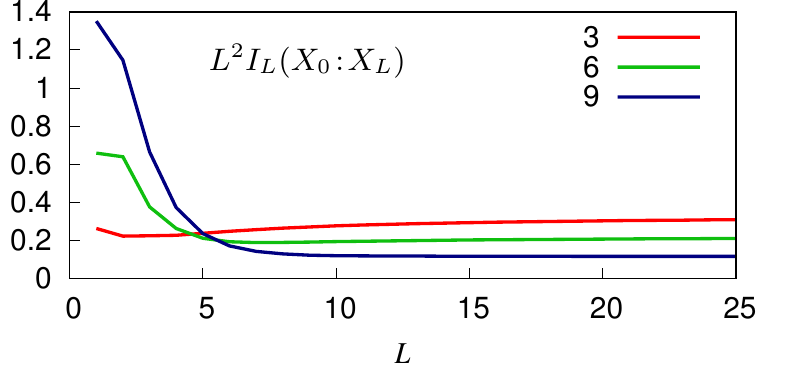}
\caption{
 $\Lt^2 I_\Lt(X_0\! :\! X_\Lt)$, the end-to-end mutual information multiplied by $\Lt^2$ for the exceptional cases of odd-$\Num$, $\Lx\in 3\Nat$, which have zero-energy modes~\cite{Nourhani+PRE-18a}.
}
\label{fig:end-to-end-mi-exceptional}  
\end{figure}

Finally, we take up the end-to-end mutual information in the anomalous cases where there are zero-energy modes, that is, when $\Num$ is odd and  $\Lx \in 3\Nat$. We shall see that it decays only as $\Lt^{-2}$.
The plots of  $\Lt^2 I_\Lt(X_0\!:\! X_\Lt)$ for the three smallest values of $\Lx$ in this class, shown in Fig.~\ref{fig:end-to-end-mi-exceptional} provide fairly convincing evidence that this is the case. A power law is perhaps not unexpected, but why not $\Lt^{-1}$?
To obtain an expression for the coefficient, the method of the previous subsection, using the Jordan canonical form of $\xfer$, would be in order. But, the $\A$-expansion is much better for illuminating the qualitative point, and we content ourselves with that.

Recall that $\Num_c$ (\ref{eq:N-critical}) is the number of modes of non-negative energy, hence including the zero-energy modes. For any configuration $X_0$ at $\tau=0$ with $\Num(X_0) \ge \Num_c = \Lx/3 + 1$, there is a finite relaxation (imaginary) time $\tau_0$, such that for $\tau_0 \ll \tau \ll \Lt$, the state is nearly certain to be in the lowest energy state of $\Num_c$ particles, which has all single-particle modes up to and including the zero-energy modes occupied. The transition to $\Num_c-2$ particle state without the zero-energy modes can take place anywhere along the length, or not at all, giving $\Lt+1$ possibilities which all have equal entropy. Essentially those are the only states of the entire cylinder in play, and the probability that the transition does not occur is proportional to $1/\Lt$.
If, on the other hand, the ``initial'' state has only $\Num_c - 2$ particles, then the transition is impossible, of course, but there is no entropy penalty for this case compared to transitioning from $\Num_c$ to $\Num_c-2$ at any particular location. By this reasoning, we arrive at the following simplified model. The number of occupied zero modes at the top of the cylinder is $N_t$ and the number at the bottom is $N_b$. They are restricted according to $N_t,N_b \in \{0,2\}$ and $N_b \le N_t$.  $N_b < N_t$ means there is a ``domain wall'' somewhere in the system. The possibilities of no domain wall and one between rings $n$ and $n+1$ for each $n$ are all equiprobable, with probability $1/L$. Joint and marginal probabilities of $N_{t}$ and $N_b$ are recorded in the following table.
\begin{center}
\begin{tabular}{c|cc||c}
\begin{picture}(27,20)
\put(1,1){$N_{b}$}    
\put(15,9){$N_{t}$}    
\put(30,-2){\line(-3,2){22}}    
  \end{picture}
 & \raisebox{6pt}{0} & \raisebox{6pt}{2} &  \\
\hline 
0 & $\;1/L$ & $1-2/L$ & $1-1/L$ \\
2 & $\;0$ & $1/L$ & $1/L$ \\
\hline \hline 
 &$\;1/L$ & $1-1/L$ & \\
\end{tabular}
\end{center}
The computation of mutual information proceeds as follows, and
involves some perhaps unexpected cancellation.
\begin{align}
I(N_b\! :\! N_t) 
= &
H(N_b) - P(N_t=0) H(N_b \!:\! N_t=0)
\nonumber \\
& - P(N_t=2) H(N_b \!:\! N_t=2) 
\nonumber \\
= &
-\left( 1 - \frac{1}{L} \right) \ln \left(1 - \frac{1}{L}\right)
-\frac{1}{L} \ln \frac{1}{L}
\nonumber \\
& +
\left( 1 - \frac{2}{L} \right)
\ln \left( \frac{1 - \frac{2}{L}}{1-\frac{1}{L}} \right)
+
\frac{1}{L} \ln \frac{\frac{1}{L}}{1-\frac{1}{L}} 
\nonumber \\
= &
-2\left( 1 - \frac{1}{L} \right)
\ln \left( 1 - \frac{1}{L} \right)
\nonumber \\
& + \left( 1 - \frac{2}{L} \right)
\ln \left( 1 - \frac{2}{L} \right)
\nonumber \\
&= L^{-2} + {\cal O}(L^{-3}).
\end{align}
The inverse-square falloff of mutual information is now explained.

\section{Conclusion}
\label{sec:conclusion}

The disorder of the zero-temperature TIAFM is similar in some way to the standard
Perron-Frobenius scenario for thermal disorder, but very different in others,
and the latter are not difficult to find. 
The fermionic formulation makes clear the origin of the abnormalities --- it is
the semi-conservation of particle number, and it facilitates detailed computations.
For a cylindrical system, mutual information is a natural tool to study the
dependence of the complex composite degrees of freedom comprised by spins in
a ring.
Zero-temperature pure phases for infinite cylinders are relatively normal
compared to thermally disordered states: the transfer matrix recovers transitivity, 
albeit on a reduced configuration space labelled by total particle number $\Num$. 
Yet the very existence of multiple phases is abnormal.
The falloff rate of the ring-to-ring mutual information is twice 
that of correlation functions, the latter equal to the spectral gap.
End-to-end mutual information on finite cylinders, by contrast, deviates
significantly from expectations for a disordered spin system in complex
dependence on circumference and periodicity or antiperiodicity. These
phenomena are, however, easily understood in terms of energies of single 
particle states in a noninteracting fermi system. 

\begin{acknowledgments}
This project was funded by the U.S. Department of Energy,
Office of Basic Energy Sciences, Materials Sciences and
Engineering Division under Grant No. DE-SC0010778, 
and by the National Science Foundation under
Grant No. DMR-1420620.
\end{acknowledgments}


%

\end{document}